\documentclass[aps,nofootinbib]{revtex4}
\usepackage{graphicx,color,amsmath}
\begin{document}
\title{Summary Report for the $e^- e^-$ session in LCWS 2004
\footnote{Plenary talk given at the International Conference on
Linear Colliders, LCWS04, 19-23 April 2004, Paris, France.}
}
\author{Kingman Cheung}
\address{
Department of Physics and NCTS, National Tsing Hua
University, Hsinchu, Taiwan, R.O.C. }
\date{\today}

\begin{abstract}
In this talk, I summarize the activities in the $e^- e^-$ session.  The 
consensus is that if the next generation $e^+ e^-$ linear 
collider wants to include an $e^- e^-$ option, the planning has to include
it as early as possible.  By doing so the extra 
cost would be only a small fraction of the total, and of course the physics
potential would be very rewarding.
\end{abstract}
\maketitle

\section{Introduction}
An electron-electron collider would be an 
ideal place to search for lepton-number
and lepton-flavor 
violation.  In fact, it is so unique that most of its physics cannot be
easily tested at other colliders, including $pp$ and $e^+ e^-$ colliders.
That is why a session in the Linear Collider Workshop was devoted to
$e^- e^-$ physics, including both accelerator technology talks and physics
talks.  In this report, I summarize the talks on both aspects.

\section{Accelerator Technology}
The first major concern is the luminosity of the $e^- e^-$ option compared to
the $e^+ e^-$ mode. Wood and Raubenheimer \cite{wood} performed luminosity
comparison for both NLC and TESLA designs due to the effects of
wakefields, disruption, and kink instability.  They found that the $e^- e^-$
option for both designs suffers more serious luminosity loss than the $e^+ e^-$
by a factor of about 10, but is recoverable to some extent with the use of
beam-based feedbacks.   Markiewicz \cite{mark} studied the IR layout,
in particular the pair-induced backgrounds.  In $e^+ e^-$ collisions,
both neutron
and charged particle backgrounds are dominated by beam-beam pairs.
The factor of a few decrease in luminosity in $e^- e^-$ option also reduces
the number of beam-beam pairs by the same factor.  Therefore, although the
$e^- e^-$ option suffers luminosity loss, the beam-beam pair backgrounds 
however is less than the $e^+ e^-$ mode.

Another important issue of the $e^- e^-$ option is the switchover in the linac,
i.e., how easy or how automatic should one expect for switching from 
$e^+ e^-$ to $e^- e^-$ and vice versa.  The goal is to obtain an optimally
functional and cost effective model for achieving the $e^- e^-$ mode.  
The requirements include: (i) quick switch, (ii) the switchover should 
cause a minimal perturbation to the running condition for $e^+ e^-$ mode,
(iii) automated means to do the switchover job.  Larsen \cite{larsen}
proposed three models for switchover.  The first one is {\it polarity
reversal model}.  Just by the name it reverses the polarity of the magnets
so that the electron and the positron can be accelerated in the same path.
The second is {\it direction reversal model}, which makes the electron to 
accelerate in the opposite direction as the positron.  The third is 
{\it independent system model}, i.e., separate beam path for positron and
electron, which is of course the most expensive.

Overall, there are quite a few issues involving in the 
$e^- e^-$ option.  However,
if one plans ahead and includes the consideration for $e^- e^-$ 
when they design the whole linear collider, the extra cost should only be 
a small
fraction of the total.  The physics paid-off would be far more than the 
cost if we plan well in advance.

\section{Physics Potential}
There were talks given by Heusch \cite{heusch}, Gunion \cite{gunion}, and
Cannoni \cite{cannoni}.  I briefly summarize their presentations.
Readers can find some nice reviews on the physics potential of 
$e^- e^-$ colliders in Refs. \cite{heusch,gunion}.  

\subsection{M\"{o}ller Scattering}
Since both electron beams can be polarized to very high purity,
one can measure the polarized cross sections, and then 
determine the 
left-right cross section asymmetries and the Weinberg angle to
very high accuracy \cite{maricano}.  One can define the following asymmetries
\begin{equation}
A_{LR}^{(1)} = \frac{ d \sigma_{LL} + d \sigma_{LR} -d \sigma_{RL}
                     -d \sigma_{RR} }
                    { d \sigma_{LL} + d \sigma_{LR} +d \sigma_{RL}
                     +d \sigma_{RR} } \,, \qquad
A_{LR}^{(2)} = \frac{ d \sigma_{LL} - d \sigma_{RR}  }
                    { d \sigma_{LL} + d \sigma_{RR}  } \;,
\end{equation}
where $d \sigma_{\alpha\beta}$ denotes the cross section for $e^-_\alpha
e^-_\beta \to e^- e^-$.
Assuming dominance by $\gamma$ and $Z$ exchanges, in the limit
$ys, (1-y)s \gg M_Z^2$, the above asymmetries can be written as
\begin{equation}
A_{LR}^{(1)} = \frac{ (1-4 s_W^2)(1+4s_W^2)}{1+ 16s_W^4 + 8(y^4 +(1-y)^4)
s_W^4 }\,,\;\;
A_{LR}^{(2)} = \frac{ (1-4 s_W^2)(1+4s_W^2)}{1+ 16s_W^4} \;.
\end{equation}
By measuring the polarized cross sections the asymmetries $A_{LR}^{(1,2)}$
can be determined and so can $s_W^2$ to high accuracy: 
$\delta s^2_W \sim 0.0003$ at $\sqrt{s}=1$ TeV and
$O(100)$ fb$^{-1}$ luminosity.  One can also make use of the $y$ dependence
on $A_{LR}^{(2)}$ to determine the $y$ dependence of $s_W^2$ \cite{maricano}.

\subsection{Lepton Flavor or Number Violation}
Certain types of new physics can be directly tested at $e^- e^-$ colliders,
in particular those involving lepton-number violating interactions.  

One unique new physics of $e^- e^-$ colliders is to search for any
doubly charged particle that has at least a weak coupling to electrons.  
Typical examples include doubly-charged Higgs bosons of some triplet Higgs
or more complicated representations \cite{hunter}, 
bilepton gauge bosons that exist in some 3-3-1 models \cite{frampton}, and
little Higgs models \cite{little}.  In Frampton's 3-3-1 model, the bilepton
gauge boson couples to the lepton triplet 
\begin{equation}
{\cal L} = \left( \ell^- \nu \ell^+ \right )^*_L \;
 \left( \begin{array}{ccc}
              & & Y^{--} \\
              & & Y^- \\
           Y^{++} & Y^{+} & \end{array} \right )\;  
 \left( \begin{array}{c}
              \ell^{-} \\
              \nu \\
              \ell^+ \end{array} \right )_L \;,
\end{equation}
where $Y$ is the new gauge boson.  $Y^{--}$ can be produced as an $s$-channel
resonance at $e^- e^-$ collisions, and then decay into $\mu^- \mu^-$.  It is
background free and $Y^{--}$ appears as a clean resonance.  In the case
of Higgs-triplet or higher Higgs representations, as long as the VEV of the
neutral component is small enough, 
it will not affect the electroweak symmetry breaking
and $\rho=1$ is preserved. In fact, triplets are desirable in a number of
neutrino mass models.   

Recently, a new class of models, called little Higgs, 
were advocated to delay the gauge hierarchy problem to 10 TeV scale.  In these
models, e.g., the littlest Higgs model \cite{little}, 
there often exists a Higgs triplet field $\Phi$, which is needed to cancel
the divergence associated with the scalar-boson loop of the Higgs boson mass.
The Higgs triplet has $T=1, Y=2$, which couples to $e^- e^-$ with a 
lepton-number violating coupling $H_{\ell\ell}^{\Phi^{--}} L \Phi L$.  
One would expect $e^- e^- \to 
\Phi^{--} \to \mu^-\mu^-, \tau^- \tau^-$ 
for lepton flavor violation.  However, 
this coupling is related by SU(2) invariance to $H_{\nu\nu}^{\Phi^0}$,
which gives a Majorana mass to left-handed neutrinos.  Therefore,
$H_{\ell\ell}^{\Phi^{--}}$ is constrained to be very small.  
Another possibility is that $\Phi^0$ develops a VEV $v'$ such that 
it couples to $WW$ and $ZZ$. Through this VEV, $\Phi^{--}$ can be produced
via $e^- e^- \to \nu\nu W^{*-} W^{*-} \to \nu \nu \Phi^{--*} \to
\nu\nu W^- W^-$.  However, it is well known that precision
electroweak measurements require $v'$ to be small \cite{gunion}.  Preliminary
estimates showed that backgrounds are too large to observe such a resonance.

\begin{figure}[t!]
\centering
\includegraphics[width=2in]{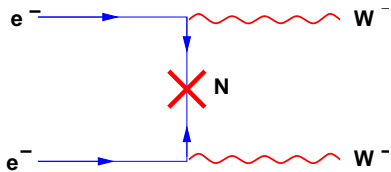}
\caption{\small Feynman diagram for the inverse neutrinoless
double beta decay via a Majorana neutrino $N$. \label{fig1}}
\end{figure}

One of the lepton-number violating reactions is the inverse of neutrinoless
double beta decay, via a $t$-channel exchange of a Majorana neutrino,
shown in Fig. \ref{fig1}.  The rate of the process is proportional 
to the square of the Majorana neutrino mass $M_N^2$.  The question is how
low the mass the collider can probe.

Another interesting process is the $e^- e^- \to \tilde{e}^- \tilde{e}^-$,
via a $t$-channel neutralino exchange \cite{feng}.  
The differential cross section of the
process is given by
\begin{equation}
\frac{d\sigma}{d \Omega} = \frac{\alpha^2 M_1^2}{2 \cos^4 \theta_{\rm w}}
\; \left ( \frac{1}{t - M_1^2} + \frac{1}{u - M_1^2} \right ) \;,
\end{equation}
where $\theta_{\rm w}$ is the Weinberg angle and $M_1$ is the bino mass.
The cross section scales as $M_1^2$, the measurement of which can determine the
soft parameter $M_1$ quite precisely.  In addition, the cross section versus
$\sqrt{s}$ is a sensitive function of $m_{\tilde{e}}$; 
in particular $e^- e^- \to  \tilde{e}^-_R \tilde{e}^-_R$
has a $S$-wave turn-on such that it provides an excellent measure of the
mass of $\tilde{e}_R^-$ at the threshold \cite{feng}.

Another related process is the slepton flavor oscillation \cite{osci}.
The word ``oscillation'' might be a little bit misleading.  It is essentially
a slepton flavor mixing that will give rise to observable lepton flavor 
violation.
In general, the matrix that diagonalizes the lepton flavors may not
diagonalize the slepton flavors, and so we could have the following
slepton mass matrix, supposed we only care the first and second generations,
\begin{equation}
{\cal M}^2_{\tilde{\ell}} = \left ( \begin{array}{cc}
                               m_{ee}^2 & m_{e\mu}^2 \\
                               m_{e\mu}^2 & m_{\mu\mu} \end{array} \right )
\end{equation}
in which the off-diagonal entry $m_{e\mu}^2$ could be of the same order
as the diagonal elements.  In this case, there will be near-maximal mixing
between selectron and smuon.  Then the following process becomes possible
\begin{equation}
e^- e^- \to \tilde{e}^- \tilde{\mu}^- \to e^- \widetilde{\chi}^0_1 \;
             \mu^- \widetilde{\chi}^0_1 \;
\end{equation}
which has one electron and one muon plus missing energies in the final
state.  The same final state could also go through 
$
e^- e^- \to \tilde{e}^- \tilde{e}^-
$, followed by $\tilde{e}^- \to \mu^- \widetilde{\chi}^0_1$ decay.

Yet, another related process, though in an entirely different framework, is
production of a pair of Kaluza-Klein electrons.  
It is in the framework of
universal extra dimensions, in which all SM particles are allowed to propagate
in extra dimensions.  In this model, there exists a KK parity assigned 
for each
SM particle and KK state such that the parity is odd for odd $n$ and even
otherwise (SM particles are even).  The lightest KK state $\gamma^{(1)}$ with
a negative KK parity is therefore stable.  In fact, it could be a potential
dark matter candidate.  The relevant process at $e^- e^-$ colliders is
$e^- e^- \to e^{-(1)} e^{-(1)}$ via a $t$-channel 
$\gamma^{(1)}$ \cite{cheng}.  
The KK electron pair will then decay into
a pair of electrons and KK photons.  Thus, the signature is two soft 
electrons plus a large missing energy.  It is unique and free from $2\gamma$
background.

So far we have only discussed 
tree-level lepton-flavor violation, Cannoni reported a work
on loop-level lepton number and flavor violation \cite{cannoni} in models
of TeV Majorana neutrino and in the supersymmetric extension of the SM.
The heavy Majorana neutrino contributes to the lepton flavor violating 
process $e^- e^- \to \mu^- \mu^-,\, \tau^- \tau^-$ through box diagrams 
\cite{cannoni}.  For Majorana neutrino masses $M_{N_i}= M_{N_j}=3$ TeV the
signal cross section can reach the level of $10^{-1}$ and $10^{-2}$ fb
for $\tau^-\tau^-$ and $\mu^-\mu^-$ channels, respectively.  The cross 
section may be too small for discovery unless the luminosity is of order
1000 fb$^{-1}$.  

In the general supersymmetric standard model (SSM), the squark and slepton mass
matrices are in general non-diagonal.  The off-diagonal matrix
elements have to be under control in order to satisfy the flavor constraints.
One usually has to adopt universal boundary conditions.  However, 
when the seesaw mechanism is embedded in the SSM, a new source of lepton-flavor
violation arises.  The seesaw requires three singlet neutrinos at 
the seesaw scale $M_R$ with additional Yukawa couplings to lepton fields:
$(Y_\nu)_{ij} H_2 N_i L_j$.
The RGE running from the GUT scale down to $M_R$ induces off-diagonal
matrix elements in $M^2_{\tilde{L}}$ given by \cite{cannoni}
\[
\left( M^2_{\tilde{L}} \right )_{ij} \sim - \frac{1}{8\pi^2} (3+a_0^2)
m_0^2 \left( Y_\nu^\dagger Y_\nu \right )_{ij} \ln 
\left( \frac{M_{GUT}}{M_R} \right )
\;,
\]
where $m_0$ is the common scalar mass and $a_0$ is related to $A_\ell$.
It in turn induces lepton flavor violation.
Cannoni {\it et al.} considered the process
$e^- e^- \to e^- \ell^- (\ell=\mu,\tau)$ involving sleptons and gauginos
in box diagrams.  In general the cross section is very small, sub-fb level,
but the 
signature is rather striking: a muon or a tau with an electron back-to-back 
in the final state.

\subsection{Strong $WW$ Scattering}
We can also study the strong $WW$ scattering at $e^- e^-$ colliders, in
particular the $W^- W^- \to W^- W^-$ mode, which has an isospin $I=2$.  
$e^- e^-$ collisions provide a unique setting for testing the isospin
$I=2$ channel.  A study was performed about 10 years ago before
the precision measurements prefer a light Higgs boson.  At any rates one
should always bear in mind that we always prepare for the surprise.  So here
I quote the table showing the event numbers \cite{wwzz}.

\begin{table}[th!]
\caption{\small \label{table1}
Cross sections (fb) for various strong $WW$ scattering models in 
$e^- e^- \to \nu_e \nu_e W^- W^-$ at $\sqrt s=2$~TeV with optimized cuts.
Those in parentheses correspond to the \# of events with hadronic $W,Z$
decays for an integrated luminosity of 300 fb$^{-1}$.}
\begin{tabular}{lccccc} 
\hline
\hline
$M_{WW}^{min}$ & SM  & Scalar & Vector   & LET & Bkgd \\
\noalign{\vskip-1ex}
& $m_H=1$ TeV & $m_S=1$ TeV & $m_V=1$ TeV & & \\
\hline
0.5 TeV
& 0.88 (130)  & 1.2 (175)  & 1.1 (167) & 1.7 (245)  &  10 (1470) \\
0.75 TeV
&  0.44 (65) & 0.72 (106) & 0.63 (93) & 1.0 (150)  & 3.5 (515)\\ 
1 TeV
& 0.15 (22)   & 0.31 (46)  & 0.26 (38) & 0.48 (71) & 1.0 (147) \\
\hline
\end{tabular}
\end{table}

Finally, the $e^- e^-$ option has unique physics potential, and the cost
for it is minimal if the linear collider project includes it as early as 
possible.

\section*{Acknowledgments}
This research was supported in part by
the National Science Council of Taiwan R.O.C. under grant no.
NSC 92-2112-M-007-053- and 93-2112-M-007-025-.

\end{document}